\documentclass{article}
\usepackage[affil-it]{authblk}
\usepackage[usenames,dvipsnames]{xcolor}
\usepackage{amsfonts}
\usepackage{amsmath,amsthm,amssymb,dsfont}
\usepackage{enumerate}
\usepackage{graphicx}	
\usepackage{subcaption}
\usepackage[margin=3cm]{geometry}
\usepackage{url}
\usepackage{todonotes}
\usepackage{bbm}
\usepackage{xcolor}
\definecolor{linkblue}{HTML}{001487}
\definecolor{ibmblue}{HTML}{FF00FF}
\usepackage{hyperref}
\hypersetup{colorlinks=true,citecolor=linkblue,linkcolor=linkblue,filecolor=linkblue,urlcolor=linkblue,breaklinks=true}

\usepackage{boldline}

\usepackage{tikz}
\usetikzlibrary{chains}
\usetikzlibrary{fit}
\usepackage{pgfplots}
\pgfplotsset{compat=1.10}
\usepgfplotslibrary{fillbetween}
\usepackage{circuitikz}
\usepackage{pifont}
\usepackage{multirow}

\usepackage{epsfig}
\usetikzlibrary{shapes.symbols,patterns} 
\usepackage{pgfplots}
\usetikzlibrary{decorations.pathmorphing}
\usetikzlibrary{decorations.pathreplacing,calligraphy}
\tikzset{snake it/.style={decorate, decoration=snake}}
\usepackage[braket, qm]{qcircuit}
\usepackage{physics}
\usepackage{enumerate}

\usepackage{mathtools}
\usepackage[nameinlink,capitalize,noabbrev]{cleveref}

\usepackage{cite}

\usepackage{mathrsfs}

\Crefname{proposition}{Prop.}{Prop.}
\crefname{proposition}{Proposition}{Propositions}

\newtheorem{theorem}{Theorem}[section]
\newtheorem*{theorem*}{Theorem}
\newtheorem{proposition}[theorem]{Proposition}

\newtheorem{lemma}[theorem]{Lemma}

\theoremstyle{remark}
\newtheorem{remark}[theorem]{Remark}

\theoremstyle{definition}

\newtheorem{example}[theorem]{Example}

\numberwithin{equation}{section}

\newcommand{\proj}[1]{\ket{#1}\!\bra{#1}}

\newcommand*{\St}{\mathrm{S}}
\newcommand*{\TPCP}{\mathrm{CPTP}}
\newcommand*{\TNCP}{\mathrm{CPTN}}
\newcommand*{\LO}{\mathrm{LO}}

\newcommand*{\LOCC}{\mathrm{LOCC}}

\newcommand*{\CNOT}{\mathrm{CNOT}}

\newcommand*{\SWAP}{\mathrm{SWAP}}

\newcommand*{\SEP}{\mathrm{SEP}}

\newcommand*{\noCC}{\underline{\mathrm{LO}}}
\newcommand*{\CC}{\underline{\mathrm{LOCC}}}

\newcommand*{\id}{\mathds{1}}

\newcommand*{\cE}{\mathcal{E}}
\newcommand*{\cF}{\mathcal{F}}
\newcommand*{\cG}{\mathcal{G}}

\newcommand*{\cI}{\mathcal{I}}

\newcommand*{\cU}{\mathcal{U}}

\newcommand{\cnott}[4]{
        \draw[#4] (#1,#2)--(#1,#3 -0.2);
        \node at (#1, #2) [circle,fill,inner sep=1pt]{};
        \draw[#4] (#1, #3) circle (0.2);
        \draw[#4] (#1 - 0.2, #3)--(#1 + 0.2,#3);
        }

\allowdisplaybreaks

\begin{document}
\title{Optimal wire cutting with classical communication}

 \author{\normalsize Lukas Brenner$^{1,2}$, Christophe Piveteau$^{2}$, and David Sutter$^{1}$}
  \affil{\small $^{1}$IBM Quantum, IBM Research Europe -- Zurich\\
  $^{2}$Institute for Theoretical Physics, ETH Zurich\\
 }
 \date{}

\maketitle

\begin{abstract}
Circuit knitting is the process of partitioning large quantum circuits into smaller subcircuits such that the result of the original circuits can be deduced by only running the subcircuits. Such techniques will be crucial for near-term and early fault-tolerant quantum computers, as the limited number of qubits is likely to be a major bottleneck for demonstrating quantum advantage. One typically distinguishes between gate cuts and wire cuts when partitioning a circuit. The cost for any circuit knitting approach scales exponentially in the number of cuts. One possibility to realize a cut is via the quasiprobability simulation technique. In fact, we argue that all existing rigorous circuit knitting techniques can be understood in this framework. Furthermore, we characterize the optimal overhead for wire cuts where the subcircuits can exchange classical information or not. We show that the optimal cost for cutting $n$ wires without and with classical communication between the subcircuits scales as $O(16^n)$ and $O(4^n)$, respectively.  
\end{abstract}

\section{Introduction}
Suppose we want to run a large quantum circuit but only have access to quantum computers with a small number of qubits.
Circuit knitting techniques allow us to determine the outcome of the circuit by cutting it into several smaller subcircuits that fit on the quantum devices, run them individually and then combine the results in a specific manner.
These techniques typically incur a sampling overhead that scales exponentially in the number of gates and wires involved in the cut.
The appeal of circuit knitting in the noisy intermediate-scale and early fault-tolerant era where the number of (logical) qubits will be limited is straightforward. However, the techniques may be useful beyond that since large-scale quantum computers will likely be built by combining multiple quantum processing units (QPUs) such that doing a large part of the calculation locally on a single QPU can be beneficial~\cite{IBM_22}.

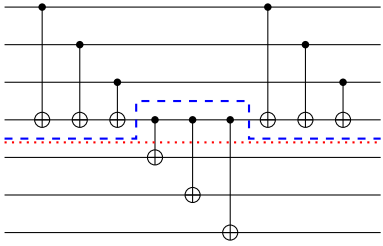
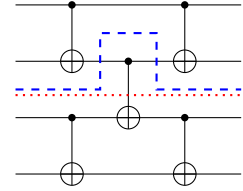
\begin{figure}[!htb]
     \centering
     \begin{subfigure}[b]{0.45\textwidth}
         \centering
    \begin{tikzpicture}[scale=0.50]
    \draw[] (0,0) -- (10,0);
    \draw[] (0,-1) -- (10,-1);
    \draw[] (0,-2) -- (10,-2);
    \draw[] (0,-3) -- (10,-3);
    \draw[] (0,-4) -- (10,-4);
    \draw[] (0,-5) -- (10,-5);
    \draw[] (0,-6) -- (10,-6);
    \cnott {1}{0}{-3}{};
    \cnott {2}{-1}{-3}{};
    \cnott {3}{-2}{-3}{};
    
    \cnott {4}{-3}{-4}{};
    \cnott {5}{-3}{-5}{};
    \cnott {6}{-3}{-6}{};
    
    \cnott {7}{0}{-3}{};
    \cnott {8}{-1}{-3}{};
    \cnott {9}{-2}{-3}{};

    \draw[dashed , thick , blue] (0,-3.5)--(3.5, -3.5) -- (3.5, -2.5) -- (6.5, -2.5) -- (6.5, -3.5) -- (10, -3.5);
    \draw[dotted, thick, red] (0,-3.6)--(10,-3.6);
    \end{tikzpicture} 
         \caption{For a decomposition into subcircuits of at most 4 qubits we need two wire cuts (dashed blue) or three gate cuts (dotted red).}
         \label{fig_line_cit}
     \end{subfigure}
     \hfill
     \begin{subfigure}[b]{0.45\textwidth}
         \centering
        \begin{tikzpicture}[scale=0.750]
        \draw[] (0,0) -- (4,0);
        \draw[] (0,-1) -- (4,-1);
        \draw[] (0,-2) -- (4,-2);
        \draw[] (0,-3) -- (4,-3);

        \cnott{1}{0}{-1}{};
        \cnott{1}{-2}{-3}{};
        \cnott{2}{-1}{-2}{};
        \cnott{3}{0}{-1}{};
        \cnott{3}{-2}{-3}{};

        \draw[dashed, thick, blue] (0,-1.5)--(1.5,-1.5)--(1.5, -.5)--(2.5, -0.5)--(2.5, -1.5) -- (4, -1.5);
        \draw[dotted, thick, red] (0,-1.6)--(4,-1.6);
    \end{tikzpicture} 
         \caption{For a decomposition into subcircuits of at most 3 qubits we need two wire cuts (dashed blue) or one gate cut (dotted red). }
         \label{fig_gate_cut}
     \end{subfigure}
        \caption{Two example circuits demonstrating the differences between wire and gate cuts.}
        \label{fig_both_cuts}
\end{figure}

Since a quantum circuit consists of wires and gates, there are two possibilities for a cut --- cutting wires or cutting gates.
\Cref{fig_both_cuts} illustrates that depending on the circuit structure either wire cuts or gate cuts can be favorable. 
Hence, for optimal circuit knitting we need to have a detailed understanding of the costs for wire as well as gate cuts.

One approach to realize gate and wire cuts is using the technique of \emph{quasiprobability simulation}, which has previously gained much interest in the fields of quantum error mitigation~\cite{TBG17,endo18,kandala19,PSBGT21,PSW22} and classical simulation algorithms~\cite{PWB15, HC17, SC19, HG19, SBHYC21}.
The idea is to probabilistically simulate the outcomes of the original quantum circuit by randomly exchanging the non-local gates across the subcircuits with local operations in a careful manner.
One can estimate the ideal expectation value of the original circuit, though with an increased variance.
This translates into a \emph{sampling overhead}, i.e., the number of shots to achieve a fixed accuracy is increased.
More precisely, suppose we can decompose a non-local gate $\cU$ as
\begin{align} \label{eq_QPD}
\cU = \sum_i a_i \cF_i \, ,
\end{align}
where $a_i$ are real numbers that may take negative values and $\cF_i$ are local operations.
During the circuit execution, the gate $\cU$ gets randomly replaced by one of the gates $\cF_i$. The sampling overhead of the this approach is given by $\kappa^2:=(\sum_i |a_i|)^2$\cite{TBG17,Piv_masterThesis}. It is thus desirable to find a decomposition in the form of~\cref{eq_QPD} that minimizes $\kappa$. 

The first circuit knitting technique was introduced by Peng \emph{et al.}\cite{PHOW20}~and focuses purely on wire cutting.
The authors phrase their technique in the language of tensor networks and they divide the tensor network describing the circuit into individual clusters which are contracted individually and combined in post-processing.
While it is not formulated in this way, the resulting procedure can in fact be seen as a special instance of quasiprobability simulation.
Later works by Mitarai and Fujii~\cite{Mitarai_2021,MF_21} first introduced gate cutting based on quasiprobability simulation for a large variety of two-qubit gates.
Another circuit knitting technique denoted \emph{entanglement forging}~\cite{forging22} uses the Schmidt decomposition of the state prepared by the larger circuit to generate a set of local circuits to sample from.
The drawback of this approach is that generally the Schmidt coefficients of a state are not known.
The authors argue that in the case of a variational circuit preparing a ground state, these Schmidt coefficients can be treated in a variational way and optimized together with the parameters of the variational circuit.

In~\cite{PS22}, we study in the context of quasiprobabilistic gate cutting whether classical communication between the smaller quantum computers (executing the subcircuits) can improve the sampling overhead of gate cutting.
We explicitly characterize the optimal sampling overhead of quasiprobabilistic gate cutting for a wide variety of gates and show that classical communication does not seem to improve the overhead for a \emph{single} instance of a gate cut.
However, we make the remarkable observation that the sampling overhead can be improved if multiple gate cuts are considered at once instead of separately.
More concretely, the optimal gate cutting technique for two $\CNOT$ gates has a strictly lower overhead than the optimal gate cutting technique for a single $\CNOT$ gate applied twice.
This idea of using classical communication to reduce the overhead of multiple cuts was later also applied to wire cuts by Lowe \emph{et al.}~\cite{xandadu_22}.
They observed that using classical communication, multiple wire cuts in \emph{parallel} could analogously be achieved cheaper than simulating all the wire cuts individually.

However, the understanding of wire cuts is still not as complete as for gate cuts.
More specifically, there is no explicit characterization of the optimal sampling overhead for wire cuts.
In this work, we fill this gap by showing that no previously known technique achieves the optimal sampling overhead in the case that the subcircuits can exchange classical information.
We construct an explicit circuit knitting technique that achieves the optimal sampling overhead.
Furthermore, we show that a speed-up for multiple wire cuts can not only be achieved when the cuts happen in parallel (as in~\cite{xandadu_22}), but also when they happen at arbitrary positions in the circuit, at the cost of a few additional ancilla qubits.

We stress that all previously proposed wire cutting techniques are an instance of quasiprobability simulation, even if they were not originally presented in that way.
Therefore, showing optimality in the setting of quasiprobabilistic circuit knitting can be considered a very general and powerful result.

\begin{figure}[!htb]
     \centering
     \begin{subfigure}[b]{0.45\textwidth}
         \centering
            \begin{tikzpicture}[scale=0.75]
                \draw[] (0,0)--(8,0);
                \draw[] (0,-1)--(8,-1);
                \draw[] (0,-2)--(8,-2);
                \draw[] (0,-3)--(8,-3);
                \cnott{1}{0}{-1}{};
                \cnott{3}{0}{-1}{};
                \cnott{2}{-1}{-2}{};
                \cnott{4}{-2}{-3}{};
                \cnott{5}{-1}{-3}{};
                \cnott{7}{-1}{-2}{};
                \draw[dashed, blue] (0, -2.5)--(3.5,-2.5)--(3.5, -0.5)--(8,-0.5);
            \end{tikzpicture}
         \caption{Parallel cut involving two wires.}
         \label{fig_parallel}
     \end{subfigure}
     \hfill
     \begin{subfigure}[b]{0.45\textwidth}
         \centering
\begin{tikzpicture}[scale = 0.75]
                \draw[] (0,0)--(6,0);
                \draw[] (0,-1)--(6,-1);
                \draw[] (0,-2)--(6,-2);
                \draw[] (0,-3)--(6,-3);
                \cnott{1}{0}{-1}{};
                \cnott{1}{-2}{-3}{};
                \cnott{2}{-1}{-2}{};
                \cnott{3}{0}{-1}{};
                \cnott{3}{-2}{-3}{};
                \cnott{4}{-1}{-2}{};
                \cnott{5}{0}{-1}{};
                \cnott{5}{-2}{-3}{};
                \draw[blue, dashed] (0,-1.5)--(1.5, -1.5)--(1.5, -0.5)--(2.5,-0.5)--(2.5, -1.5)--(3.5, -1.5)--(3.5, -0.5)--(4.5, -0.5)--(4.5, -1.5)--(6, -1.5);
            \end{tikzpicture}
         \caption{Arbitrary wire cut involving four qubits. }
         \label{fig_arbitrary}
     \end{subfigure}
        \caption{Two examples explaining the settings~\eqref{it_b} and~\eqref{it_c}.}
        \label{fig_settings}
\end{figure}
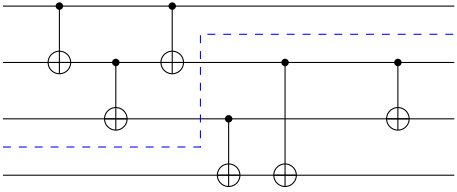
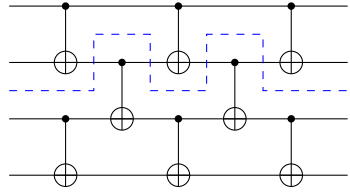

In the simplest scenario, we only cut one single wire in the whole circuit.
Once the circuit contains multiple wire cuts, things get more tricky. As observed in previous works on circuit knitting\cite{PS22,xandadu_22} the sampling overhead of doing multiple cuts can be lower than applying the optimal cutting procedure to individual cuts separately.
More concretely, Lowe \emph{et al.}\cite{xandadu_22} proposed a technique that outperforms individual optimal wire cuts when the cuts are performed in parallel, i.e., in the same time slice of the circuit.
Therefore, when talking about the overhead of cutting multiple wires, we distinguish between two different scenarios (visualized by~\cref{fig_settings}) for which we aim to characterize the optimal cost in terms of sampling overhead.
In each scenario we cut $n$ wires.
\begin{enumerate}[(i)] 
\item \label{it_b} \textbf{Parallel wire cuts:} We consider $n$ wires that are cut in parallel, which means in the same time slice of the circuit.
\item \label{it_c}\textbf{Arbitrary wire cuts:} The $n$ wires that are cut lie at arbitrary positions in the circuit.
\end{enumerate}

Note that a circuit with a single wire cut can be thought as the $n=1$ limit of either one of these scenarios.
Since parallel cuts are a special instance of arbitrary cuts, they clearly cannot involve a higher sampling overhead.
For each of the two scenarios mentioned above, we study whether allowing the two involved subcircuits to exchange classical communication affects the optimal sampling overhead.
Clearly, the sampling overhead cannot increase if we allow for classical communication.
We aim to quantify to what extend classical communication can help to improve the sampling overhead in the two scenarios.
Some practical considerations about the implementation of classical communication are discussed in~\cref{sec_discussion}.

\begin{table}[!htb]
\centering
\bgroup
\def\arraystretch{1.5}
  \begin{tabular}{V{2.5}lV{2.5}c|cV{2.5}c|cV{2.5}}
  \clineB{2-5}{2.5}  
\multicolumn{1}{cV{2.5}}{}&
      \multicolumn{2}{cV{2.5}}{\!\textbf{Without classical comm.}\!} &
      \multicolumn{2}{cV{2.5}}{\textbf{With classical communication}} \\ \cline{2-5}
\multicolumn{1}{cV{2.5}}{}& \!\!Best known\!\! & Optimal & Best known & Optimal  \\ 
\clineB{1-5}{2.5}
\!\!\eqref{it_b} Parallel & $4^{2n}$~\cite{PHOW20} & $4^{2n}$ [\Cref{prop:noCC_parallel}] & \!\!$\min\{4^{2n},(2^{n+1}+1)^2\}$~\cite{xandadu_22}\!\! & $(2^{n+1}-1)^2$ [\Cref{prop:optimal_CC}]  \\
\clineB{1-5}{2.5}
\!\!\eqref{it_c} Arbitrary & $4^{2n}$~\cite{PHOW20} & $4^{2n}$ [\Cref{prop:noCC_parallel}] & $4^{2n}$~\cite{PHOW20} & \!$(2^{n+1}-1)^2$ [\Cref{sec_arbitrary_cuts_LOCC}]\!  \\
\clineB{1-5}{2.5}
  \end{tabular}
\egroup
\caption{Overview of sampling overhead for cutting $n$ wires in the parallel or arbitrary scenario (explained in the main text) with or without classical communication. 
The ``best known" column indicates the sampling overhead of the best algorithm prior this work.
The ``optimal" column denotes the optimal sampling overhead and the corresponding reference indicates where the proof of optimality can be found in this paper.}
		\label{tab_results}
\end{table}

\paragraph{Results}
\cref{tab_results} shows the main results of this paper.
Recall that the sampling overhead is given by $\kappa^2$ where $\kappa$ is the one-norm of the coefficients of the quasiprobability decomposition (see~\cref{eq_QPD}).
We characterize the optimal sampling overheads for the two scenarios~\eqref{it_b} and~\eqref{it_c} with and without the help of classical communication between the cut subcircuits. For some settings we considerably improve on the best previously known methods, for other settings we show that the best existing technique is optimal and hence cannot be further improved.

We would like to emphasize three points from~\cref{tab_results}. First, we observe that for single wire cuts ($n=1$) classical communication can already strictly improve the optimal sampling overhead from $16$ to $9$.
Such a provable separation between the two cases with and without classical communication is remarkable, because gate cutting typically does not exhibit such a separation~\cite[Corollary~4.4]{PS22}.
Second, in analogy with gate cutting, wire cuts with classical communication feature a strict submultiplicativity property in the sense that two separate wire cuts are more expensive than two wire cuts that are considered together.
This strict submultiplicativity is not present when classical communication is not allowed.
While a submultiplicative (though suboptimal) behavior was previously known for scenario~\eqref{it_b} with classical communication~\cite{xandadu_22}, we show for the first time that it can also be observed for arbitrary cuts.
Finally and third, we see that the ability to exchange classical information between the subcircuits allows us to considerably reduce the sampling overhead in setting~\eqref{it_b} as well as setting~\eqref{it_c}.

\section{Preliminaries}
In this section we review the quasiprobability simulation technique which will later be used to perform wire cutting.
Furthermore, we introduce some results on the quasiprobabilistic simulation of non-local gates and non-local states that will play an important role for the optimality analysis of wire cutting.

\subsection{Notation}
For finite dimensional Hilbert spaces $A$ and $B$, let $\mathrm{L}(A,B)$ denote the set of linear maps that map from $A$ to $B$.
We use the short-hand notation $\mathrm{L}(A)\coloneqq \mathrm{L}(A,A)$ and thus the set of superoperators acting on the system $A$ is given by $\mathrm{L}(\mathrm{L}(A))$.
We denote by $\TPCP(A)$ the subset of $\mathrm{L}(\mathrm{L}(A))$ of superoperators that are trace-preserving and completely positive.
Similarly, we denote by $\TNCP(A)$ the subset of $\mathrm{L}(\mathrm{L}(A))$ of trace non-increasing completely positive maps.
The set of Hermitian operators on $A$ is denoted by $\mathrm{H}(A)$ and the set of density operators on $A$ is denoted by $\St(A)$.
Moreover, we write $\|\cdot \|$ for the operator norm and $\|\cdot\|_1$ for the trace norm.

\subsection{Quasiprobability simulation}\label{sec:qpsim}
Consider a quantum computer and denote the system of its qubits by $A$.
Furthermore, suppose the computer is not universal, i.e., there exists some unitary $U$ on $A$ that the quantum computer cannot execute.
Put differently, the channel $\mathcal{U}$ induced by the gate $U$ lies outside the set of superoperators $S\subset \mathrm{L}(\mathrm{L}(A))$ that can be achieved by the computer.
The goal of the quasiprobability simulation technique is to simulate the execution of the gate $U$ while only having access to the operations that lie in the set $S$.
Quasiprobability simulation can recover the expected value of the measurement outcomes of the circuit at the cost of an additional sampling overhead.
This kind of setup occurs in multiple fields in near-term quantum computing such as error mitigation~\cite{TBG17,endo18,kandala19,PSBGT21,PSW22} and classical simulation of near-Clifford circuits~\cite{PWB15, HC17, SC19, HG19, SBHYC21}.
For the purposes of error mitigation, $S$ describes the noisy operations (gates, measurements and combinations thereof) that a faulty near-term quantum computer could realize.
In the case of classical simulations, $S$ is chosen to be the operations realizable by Clifford gates and measurements in the computational basis (which are efficiently classically simulatable due to the Gottesman-Knill theorem) and $U$ would be some non-Clifford gate like the $T$-gate.
Recently, it was realized that gate cutting can also be cast in the picture of quasiprobability simulation~\cite{Mitarai_2021,MF_21,PS22}, by choosing $U$ to be a non-local gate acting across the cut parts of the quantum circuit and $S$ to be operations that only act locally in some sense.
The exact procedure of applying quasiprobability simulation to gate cutting will be elaborated upon in~\cref{sec:qp_gate_cutting}.

The central ingredient for the quasiprobability simulation is a \emph{quasiprobability decomposition} (QPD) as shown in~\cref{eq_QPD} of the channel $\mathcal{U}$ into some operations $\mathcal{F}_i\in S$.
Consider as an illustrative example a simple quantum circuit that initializes a state $\rho$, performs the unitary $U$ on it and then measures the state according to some observable $O$.
The ideal expectation value of the circuit outcome that we want to obtain is thus given by $\tr[O\, \mathcal{U}(\rho)]$.
Inserting the QPD from~\cref{eq_QPD} yields
\begin{align}\label{eq_exp_value}
    \tr[O \, \mathcal{U}(\rho)] = \sum_i p_i \tr[O \, \mathcal{F}_i(\rho)] \kappa \, \mathrm{sign}(a_i) \, ,
\end{align}
where $\kappa\coloneqq\sum_i|a_i|$ and $p_i\coloneqq|a_i|/\kappa$ is a probability distribution.
\cref{eq_exp_value} gives us a Monte Carlo approach to estimate the expectation value of the original circuit: For each shot of the circuit, we randomly replace the unachievable gate $\mathcal{U}$ with one of the achievable gates $\mathcal{F}_i$ with probability $p_i$ and weight the outcome with $\kappa\,\mathrm{sign}(a_i)$.
This method gives us an unbiased estimator for the desired quantity, at the cost of an increased variance.
One can verify that the number of shots to estimate the expectation value to some fixed accuracy increases by a factor proportional to $\kappa^2$~\cite{TBG17,endo18,Piv_masterThesis}. This justifies to call $\kappa^2$ the sampling overhead.

To minimize the sampling overhead it is crucial to find an optimal QDP that minimizes the one-norm of the coefficients $a_i$.
This has motivated the definition of the \emph{$\gamma$-factor} of the channel $\mathcal{U}$ with respect to a set $S$ as
\begin{equation} \label{gamma QPD}
    \gamma_S(\mathcal{U}) := \min \left\{ \sum_{i=1}^m |a_i|: m\geq 1, \mathcal{U} = \sum_{i=1}^m a_i \mathcal{F}_i, \,  a_i \in \mathbb{R}, \, \mathcal{F}_i \in S \right\} \, ,
\end{equation}
which exactly captures the optimal sampling overhead required to simulate $\cU$ using the operations in $S$.
The intuition here is the following: When $\mathcal{U}$ is close to $S$, then $\gamma_S(\mathcal{U})$ is almost $1$ and thus the sampling overhead is small.
The farther away $\mathcal{U}$ is from $S$, the more costly it becomes to simulate $\mathcal{U}$ using operations in $S$.
For more details on the quasiprobability simulation method, we refer the reader to~\cite{TBG17,endo18,Piv_masterThesis}. Note that depending on $\cU$ and $S$,~\cref{gamma QPD} is a complicated optimization problem that is unclear how to be solved.

In practice, one is mostly interested in applying this procedure to circuits that contain multiple unreachable gates $(U_i)_{i=1}^n$ that lie outside of $S$.
For each one of these gates, one obtains an associated QPD with one-norm of the quasiprobability coefficients $\kappa_i$.
One can easily convince oneself that these QPDs of individual gates can be combined into a QPD for the complete circuit, leading to a total sampling overhead of $\prod_{i=1}^n\kappa_i^2$ for simulating the complete circuit.
This implies that using the individual optimal QPD for each gate leads to a simulating overhead of $\prod_{i=1}^n\gamma_S(\mathcal{U}_i)^2$ for the whole circuit.
However, as we will see in~\cref{sec:qp_gate_cutting}, there are instances where combining optimal QPDs for the individual gates does not result into an optimal QPD for the complete circuit, i.e., local optimality does not necessarily imply global optimality.

\subsection{Trace-nonincreasing and non-positive operations}\label{sec:tni_and_np_ops}
A priori, one could reasonably expect the set of achievable operations $S$ has to be a subset of trace-preserving completely positive maps $S\subset\TPCP(A)$ since they precisely constitute the set of all physically realizable quantum channels.
However, it was realized in~\cite{endo18} that it can be useful (or often even necessary) to allow $S$ to also include trace-nonincreasing maps, i.e., $S\subset\TNCP(A)$.
This can be done because any trace-nonincreasing map can be effectively simulated by some measurement process and post-selection of the corresponding measurement outcome.
More precisely, any map $\mathcal{E}\in\TNCP(A)$ can be extended to a trace-preserving map: $\exists \mathcal{F}\in\TNCP(A)$ s.t. $\mathcal{E}+\mathcal{F}\in\TPCP(A)$.
To simulate $\mathcal{E}$, one can perform the trace-preserving completely positive map
\begin{align}
    \rho_A \mapsto \mathcal{E}(\rho)_A\otimes\ketbra{0}{0}_E + \mathcal{F}(\rho)_A\otimes\ketbra{1}{1}_E \, ,
\end{align}
where $E$ is a qubit system.
One then measures $E$ in the computational basis and postselects for the outcome $0$.
In practice, this is done by multiplying the final outcome of the circuit by $0$ in case the measurement outcome $1$ is obtained.

This trick was later generalized by Mitarai and Fujii~\cite{Mitarai_2021,MF_21} who realized that one can even allow $S$ to contain non-completely positive maps, as long as it can be written as a difference of completely-positive trace-nonincreasing maps that add up to another completely-positive trace-nonincreasing map:
\begin{align}
  S \subset \mathrm{D}(A) := \{\mathcal{E}\in \mathrm{L}(\mathrm{L}(A)) | \, \exists \mathcal{E}^+,\mathcal{E}^-\in \TNCP(A) : \mathcal{E}=\mathcal{E}^+-\mathcal{E}^-,\, \mathcal{E}^++\mathcal{E}^-\in \TNCP(A)\} \, .
\end{align}
The idea here is very similar: any such map $\mathcal{E}=\mathcal{E}^+-\mathcal{E}^-\in \mathrm{D}(A)$ can be simulated using the trace-nonincreasing completely positive map
\begin{align}
    \rho_A \mapsto \mathcal{E}^+(\rho)_A\otimes\ketbra{0}{0}_E + \mathcal{E}^-(\rho)_A\otimes\ketbra{1}{1}_E \, ,
\end{align}
measuring the qubit $E$ in the computational basis and and correspondingly weighting the final measurement outcome of the circuit by $+1$ or $-1$ depending on the outcome.

\subsection{Gate cutting and state preparation via quasiprobability simulation}\label{sec:qp_gate_cutting}
Quasiprobability simulation can be straightforwardly applied to simulating a nonlocal gate with only local operations.
Consider a non-local unitary $U_{AB}$ acting on the joint Hilbert space $A\otimes B$.
We want to cut this gate across the separation of $A$ and $B$, i.e., we want to simulate that unitary using only operations that act locally on $A$ and $B$.
If we do not allow the two circuit parts to communicate classically, this translates to a set $S=\LO(A,B)$ where
\begin{equation}
    \LO(A,B) := \{ \mathcal{A}\otimes \mathcal{B} | \mathcal{A}\in \mathrm{D}(A), \mathcal{B}\in \mathrm{D}(B)\} \, .
\end{equation}
If instead we allow the two circuit parts to communicate classically, we consider a larger set $S=\LOCC(A,B)$ where $\LOCC(A,B)$ consists of all protocols containing operations in $\LO(A,B)$ as well as classical communication between $A$ and $B$.\footnote{Note that the precise definition of $\LOCC$ is notoriously complicated due to the possibly unbounded number of classical communication rounds and the fact that later local operations can in general depend on all the previous communication. The interested reader may consult~\cite{CLMOW14} for more details.}
The optimal sampling overhead of simulating $U$ with and without classical communication is thus given by $\gamma_{\LO}(\mathcal{U})$ and $\gamma_{\LOCC}(\mathcal{U})$.

In~\cite{PS22}, we characterized the optimal overheads with and without classical communication for all Clifford gates as well as a large class of two-qubit unitaries.
For example, we showed that
\begin{align} \label{eq_CNOT_cost}
\gamma_{\LO}(\CNOT)=\gamma_{\LOCC}(\CNOT)=3 \qquad \textnormal{and} \qquad \gamma_{\LO}(\SWAP)=\gamma_{\LOCC}(\SWAP)=7 \, .
\end{align}
Therefore, classical communication does not help for the task of simulating a single $\CNOT$ or $\SWAP$ gate.
Surprisingly, we also showed that the situation changes drastically if one considers the cutting of multiple $\CNOT$ and $\SWAP$ gates and we showed that a significantly lower sampling overhead can be achieved by not considering every instance of a non-local gate separately but instead by treating them in a joint manner.

The key ingredient in that regard is to consider the optimal overhead required to simulate the preparation of some arbitrary state $\rho_{AB}$ on $A\otimes B$, which we also denote with the symbol $\gamma$:
\begin{equation}
     \gamma_S(\rho_{AB}) = \min \left\{ \gamma_S(\mathcal{E}): \mathcal{E}\in \mathrm{L}(\mathrm{L}(A\otimes B)) \text{ s.t. } \mathcal{E}(\proj{0}) = \rho_{AB} \right\} \, .
\end{equation}
where $\ket{0}$ is some fixed separable state, which is typically chosen as the all-zero state of the involved qubits. 
The following lemma shows that for quasiprobabilistic state preparation, there is also no advantage in exploiting classical communication and it shows that the $\gamma$-factor can be regarded as the optimal decomposition into separable states.
\begin{lemma}[{\!\!\cite{PS22}}]
For any bipartite state $\rho_{AB}$ we have
    \begin{equation}
        \gamma_{\LO}(\rho_{AB}) =  \gamma_{\LOCC}(\rho_{AB})
    \end{equation} 
and $\gamma_{\LOCC}(\rho_{AB}) = \min \left\{ a_+ + a_-: \rho_{AB} = a_+ \rho_+ - a_- \rho_-, \ \rho_\pm \in \SEP(A,B), \ a_\pm \geq 0 \right\}$.
\end{lemma}
Here, $\SEP(A,B)$ denotes the set of separable states on $A\otimes B$.
Moreover, using tools from the resource theory of entanglement, the $\gamma$-factor for an arbitrary pure state can be explicitly computed in terms of its Schmidt coefficients
\begin{lemma}[{\!\!\cite{PS22}}]
Let $\ket{\psi}_{AB}$ be a bipartite state with Schmidt coefficients $\{\alpha_i\}_i$. Then
    \begin{equation}
        \gamma_{\LOCC}(\ketbra{\psi}{\psi}_{AB}) = 2\left(\sum_i \alpha_i\right)^2 - 1 \, .
    \end{equation} 
\end{lemma}
Most importantly, the $\gamma$-factor of a state behaves strictly submultiplicatively.
For instance, the $\gamma$-factor for a Bell pair shared across $A$ and $B$ is given by $\gamma_{\LOCC}(\text{1 Bell pair})=3$ whereas the $\gamma$-factor for $n$ shared Bell pairs across $A$ and $B$ is given by
\begin{equation}\label{eq:gamma_n_bell_pairs}
    \gamma_{\LOCC}(\text{n Bell pairs})=2^{n+1}-1 < 3^n \, .
\end{equation}
Hence, in the asymptotic limit of $n\rightarrow\infty$, the effective cost of simulating the preparation of a Bell pair tends to $\lim_{n\rightarrow\infty} (2^{n+1}-1)^{1/n} = 2$.\footnote{Recall from~\cref{sec:qpsim} that the total overhead of simulating multiple gates behaves multiplicatively in the individual $\gamma$-factors.}

The main idea of the protocol in~\cite{PS22} is to reduce the realization of a Clifford gate to the preparation of some shared entangled quantum state through the gate teleportation protocol.
For instance, a $\CNOT$ gate can be realized using a shared Bell pair and $\LOCC$ operations.
Using the strict submultiplicativity of the $\gamma$-factor then allows us to simulate $n$ $\CNOT$ gates more cheaply than optimally simulating $n$ $\CNOT$ gates individually.
The need for classical communication in this protocol stems from the gate teleportation protocol, which involves bidirectional exchange of classical information between $A$ and $B$.

\section{Relating wire cutting to gate cutting}\label{sec:reduction_wire_to_gate}
The core idea of wire cutting using the quasiprobability simulation framework is to find a QPD of the identity channel into some set $S$ where $S$ captures whether the two quantum computers allow for classical communication or not.
However, in contrary to gate cutting, it is a priori less clear how to mathematically model the restriction of local operations with and without classical communication for the task of wire cutting.
In the following, we model the process of wire cutting in terms of gate cutting, which allows us to find a characterization of the optimal sampling overhead.
This framework for modelling wire cutting is very powerful and we argue in~\cref{app_QPD_methods} that it covers the methods from all previously proposed techniques for wire cutting .

\begin{figure}[h]
    \centering
    \begin{tikzpicture}
        \draw[thick] (0,0) -- (2,0);
        \draw[dashed, thick ,red ] (1,0.5)--(1,-0.5);

        \draw[thick] (4.0,0.5)--(6,0.5);
        \draw[thick] (5.6,-0.5)--(6,-0.5);
        \draw[thick] (7,-0.5)--(8.5,-0.5);
        \draw[thick] (6,1)--(7,1)--(7,-1)--(6,-1)--(6,1);
        \draw (7,0.5)--(7.55,0.5);
        \draw (7.6, 0.65)--(7.6, 0.35);
        \draw (7.65, 0.6)--(7.65, 0.4);
        \draw (7.7, 0.55)--(7.7, 0.45);
        \node at (6.5,0) {$\Phi$};
        \node at (5.2, -0.5) {$\proj{0}$};
        \node at (5.7, 0.8) {$A$};
        \node at (5.7, -.8) {$B$};
        \draw[dotted, blue, thick] (4.7, 1.5)--(8,1.5)--(8,-1.5)--(4.7,-1.5)--(4.7,1.5);
        \node at (6.5, 1.8) {\color{blue}$\mathcal{I}_{A\rightarrow B}$};
        \node at (3,0) {$\simeq$};

    \end{tikzpicture}
    \caption{The task of wire cutting can be translated into the framework of quasiprobabilistic gate cutting. This figure is explained in the main text.}
    \label{fig_teleportation}
\end{figure}
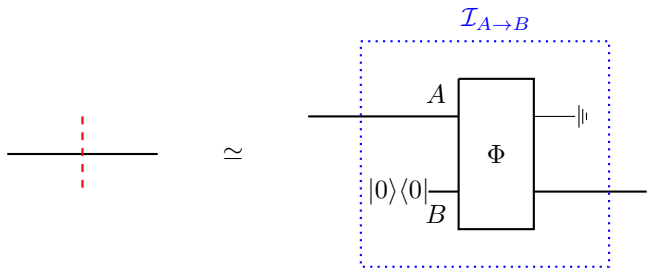

Consider a wire cut that divides a circuit into two parts which are to be executed on two separate quantum computers, as depicted in the left-hand side of~\cref{fig_teleportation}.
Wire cutting can be thought as quasiprobabilistically simulating the transmission of the qubit from one computer to the other.
Both computers clearly require a physical qubit system to store the qubit before or after the wire cut respectively---we denote these systems as $A$ and $B$.
Without loss of generality, one can assume that the initial state of the ``receiving" system $B$ is the $\ket{0}$ state.
The process of wire cutting can thus be regarded as the quasiprobabilistic simulation of some non-local operation $\Phi\in \mathrm{L}(\mathrm{L}(A\otimes B))$ which acts as the identity $\mathcal{I}_{A\rightarrow B}$ when the system $A$ is traced out afterwards:
\begin{equation}\label{eq:phiA_constraint}
    \forall X\in \mathrm{L}(A): \tr_A\left[ \Phi(X\otimes\ketbra{0}{0}_B) \right] = X \, .
\end{equation}
The optimal simulation overhead of the wire cut is thus characterized by the smallest achievable $\gamma_S(\Phi)$ over all $\Phi$ satisfying~\cref{eq:phiA_constraint}, where we either choose $S=\LOCC(A,B)$ or $S=\LO(A,B)$ depending on whether we allow for classical communication between the systems or not.

We now mathematically formalize the notions outlined above to allow for more precise statements.
Let $A$ be a finite-dimensional quantum system and let $B$ be a copy of it.
For $\cE \in \mathrm{L}(\mathrm{L}(A\otimes B))$ we define the map $\Gamma: \, \mathrm{L}(\mathrm{L}(A \otimes B)) \to \mathrm{L}(\mathrm{L}(A), \mathrm{L}(B))$ by
\begin{equation} \label{Gamma map}
    \Gamma \left[ \mathcal{E} \right](X_A) = \tr_A \left[ \mathcal{E}(X_A \otimes \proj{0}_B) \right] \, \text{ for } X_A\in \mathrm{L}(A) \, .
\end{equation}
We denote the images of $\LO(A,B)$ and $\LOCC(A,B)$ under this map by 
\begin{align} \label{eq_noCC_CC}
    \noCC(A,B) := \Gamma \left[ \mathrm{LO(A,B)} \right] \qquad \textnormal{and} \qquad \CC(A,B) := \Gamma \left[ \mathrm{LOCC(A,B)} \right] \, .
\end{align}

The following lemma shows that the optimal sampling overhead of a wire cut is given by some $\gamma$-factor of the identity channel.
\begin{lemma} \label{same gamma factors} Let $S \in \{\mathrm{LO}(A,B), \mathrm{LOCC}(A,B) \}$ and $\mathcal{F} \in \mathrm{L}(\mathrm{L}(A),\mathrm{L}(B))$. Then
\begin{equation} \label{eq_lemma_gamma}
    \min_{\Phi_{\mathcal{F}}} \left\{ \gamma_{S}(\Phi_{\mathcal{F}}) : \Phi_{\mathcal{F}} \in \mathrm{L}(\mathrm{L}(A\otimes B)), \Gamma(\Phi_{\mathcal{F}})=\cF  \right\}  = \gamma_{\Gamma[S]}(\mathcal{F})\, .
\end{equation}
\end{lemma}
Note that in the context of wire cutting we are mostly interested in the case $\mathcal{F}=\cI_{A\rightarrow B}$.
This lemma is useful, because instead of having to consider QPDs of all possible $\Phi$ fulfilling~\cref{eq:phiA_constraint}, we only need to think about QPDs of the identity channel over either the set $\CC(A,B)$ or $\noCC(A,B)$.
\begin{proof}
    We start off by showing that the expression on the right-hand side of~\cref{eq_lemma_gamma} is well-defined, i.e. that there is a QPD that achieves $\gamma_{\Gamma[S]}(\cF)$. First, note that
    $\Gamma[S]$ must be compact, because $S$ is compact\footnote{Similar to~\cite{PS22} we restrict ourselves to $\LOCC$ protocols with a bounded number of rounds, where the bound can be chosen arbitrarily large.} and $\Gamma$ is continuous as it is linear.
    As a result, we can apply the argumentation from~\cite[Appendix~A]{PS22} to see that $\gamma_{\Gamma[S]}(\mathcal{F})$ is well defined.
    As is outlined further below in the proof, any QPD of $\mathcal{F}$ in the set $\Gamma[S]$ induces a map $\Phi_{\mathcal{F}}$ with a QPD with identical coefficients. Therefore, the fact that the minimum of the right-hand side is achieved implies that the minimum of the left-hand side is also achieved, so the left-hand side is also well-defined.
    
    Consider a map $\Phi_{\mathcal{F}}$ that achieves the minimum of the left-hand side as well as an optimal QPD
    \begin{equation}
        \Phi_{\mathcal{F}} = \sum_{i} a_i \mathcal{E}_i \quad \textnormal{with} \quad \mathcal{E}_i \in S\, ,
    \end{equation}
    such that $\sum_{i}|a_i| = \gamma_{S}(\Phi_{\mathcal{F}})$. By linearity of the map $\Gamma$ we find
    \begin{equation}
        \mathcal{F} = \Gamma\left[ \Phi_{\mathcal{F}} \right] = \sum_i a_i \Gamma \left[ \mathcal{E}_i \right] \, ,
    \end{equation}
which implies $\gamma_{\Gamma[S]}(\mathcal{F}) \leq \gamma_{S}(\Phi_{\mathcal{F}})$.
To show the other direction, we will find a valid map $\Phi_\mathcal{F}$ which has the same $\gamma$-factor as $\cF$. To do so, we start with an optimal QPD of $\mathcal{F}$
\begin{equation}
        \mathcal{F} = \sum_{i} b_i \mathcal{G}_i \quad \textnormal{with} \quad \mathcal{G}_i \in \Gamma\left[ S\right] \quad \textnormal{and} \quad \sum_i |b_i|=\gamma_{\Gamma[S]}(\mathcal{F}) \, ,
\end{equation} 
which exists due to our previous reasoning. We can construct a valid map $\Phi_{\cF}$ the following way: For each $i$ pick an element $\cE_i \in \Gamma^{-1}[\{\cG_i\}]$ (not necessarily unique) and set $\Phi_{\cF} = \sum_i b_i  \cE_i$ such that we have $\Gamma[\Phi_{\cF}] = \sum_i b_i  \mathcal{G}_i = \cF$ and so $\gamma_{\Gamma[S]}(\mathcal{F}) = \gamma_{S}(\Phi_{\mathcal{F}})$ by construction.
\end{proof}

In the following, we provide a simple characterization of the set $\noCC(A,B)$.
\begin{lemma} \label{lem:equivalent_set}
   We have
\begin{align}
&\noCC(A,B) \nonumber \\ 
&=  \left\{ \! \mathcal{E}\! \in \!\mathrm{L}( \mathrm{L}(A))\!:  \mathcal{E}(X) \!=\! \sum_{j} b_j \tr[\Lambda_j X]\rho, \, \Lambda_j \geq 0, \, \sum_{j} \Lambda_j \leq \id, \, b_j = \pm 1, \, \rho\in\mathrm{H}(A), \|\rho\|_1\leq 1 \right\} \label{eq_first_ds} \\
&=\left\{ \mathcal{E} \in \mathrm{L}( \mathrm{L}(A)): \, \mathcal{E}(X) = \tr[O X]\rho, \, O={O}^\dagger, \, \|O \| \leq 1 , \, \rho\in\mathrm{H}(A), \|\rho\|_1\leq 1 \right\} \, . \label{eq_second_ds}
\end{align}
\end{lemma}
The statement of this lemma has an intuitive explanation: $\noCC(A,B)$ consists of maps that describe the measurement of some positive operator-valued measure (POVM) $\{\Lambda\}_j$ on the system $A$ followed by the preparation of the state $\rho$ on system $B$.
Since no classical communication between $A$ and $B$ is allowed, the state $\rho$ may not depend on measurement result on $A$.
However, in accordance to the discussion of non-positive operations in~\cref{sec:tni_and_np_ops}, the final output of the circuit may be weighted be either $+1$ or $-1$ depending on the measurement outcome of the POVM, which is captured by the coefficients $b_j\in \{\pm 1\}$.
\begin{proof}
We start by showing~\cref{eq_first_ds}.
Pick an arbitrary element in $\LO$ of the form $\left(\mathcal{E}^{+} - \mathcal{E}^{-} \right) \otimes \mathcal{F}$ where $\mathcal{E}^{\pm}\in\TNCP(A)$ and $\mathcal{F}\in \mathrm{D}(B)$.
Our goal is to show that the image under $\Gamma$ lies in the set on the right-hand side of~\cref{eq_first_ds}.
Note that for $\mathcal{E}^{\pm}$ there exist Kraus representations $\{{K_i}^{+} \}_i$ and $\{{K_j}^{-} \}_j$ such that 
        \begin{align}
            \sum_{i} {\left({K_i}^{+}\right)}^\dagger {K_i}^{+} + \sum_{j} {\left({K_j}^{-}\right)}^\dagger {K_j}^{-} \leq \id \, .
        \end{align}
        We compute the action of the map $\Gamma$
        \begin{align}
            \Gamma[ \left(\mathcal{E}^{+} - \mathcal{E}^{-} \right) \otimes \mathcal{F}] (X) &= \mathrm{tr}\left[  \left(\mathcal{E}^{+} - \mathcal{E}^{-} \right) (X) \right] \mathcal{F}(\proj{0}) \\
            &= \tr[ \left(\sum_{i} {\left({K_i}^{+}\right)}^\dagger {K_i}^{+} - \sum_{j} {\left({K_j}^{-}\right)}^\dagger {K_j}^{-}\right)X]\mathcal{F}(\proj{0}) \\
            &= \sum_{k} b_k \tr[\Lambda_k X] \mathcal{F}(\proj{0}) \, ,
        \end{align}
        with $\{\Lambda_k\}_k = \{ {\left({K_i}^{+}\right)}^\dagger {K_i}^{+} \}_i \cup \{ {\left({K_j}^{-}\right)}^\dagger {K_j}^{-} \}_j$ and $b_k = \pm 1$.
        Furthermore, by definition of $\mathrm{D}(A)$, we can write $\mathcal{F}(\proj{0})=\rho_+ - \rho_-$ s.t.\ $\rho_\pm \geq 0$ and $\tr[\rho_+ + \rho_-] \leq 0$.
        This implies $\|\rho\|_1\leq \|\rho_+\|_1 + \|\rho_-\|_1 = \tr[\rho_+] + \tr[\rho_-] = 1$.
        
        To check the converse statement, we have to show that for every map in the set on the right-hand side of \cref{eq_first_ds}, we can find a local operation which is mapped to it under $\Gamma$.
        To see this, for a given channel $\cG(X) = \sum_{j} b_j \tr[\Lambda_j X]\rho$ one can pick any local operation $\cF \in \mathrm{D}(B)$ that fulfills $\cF(\proj{0}) = \rho$.
        To see that such a $\cF$ must exist, one can separate $\rho$ into
        \begin{align}
            \rho=\rho_+-\rho_- \text{ where } \rho_+ = \sum_{i:\lambda_i\geq 0}\lambda_i\proj{\lambda_i}, \quad \rho_- = \sum_{i:\lambda_i< 0}-\lambda_i\proj{\lambda_i}, 
        \end{align}
        and $\lambda_i,\ket{\lambda_i}$ denote the eigenvalues and eigenstates of $\rho$.
        Moreover, due to $\Lambda_j$ being positive, we can always find operators $K_j \in \mathrm{L}(A)$ such that $\Lambda_j = K_j^\dagger K_j$. Now, set 
\begin{equation}
    \cE^\pm(X) := \sum_{j: b_j=\pm 1} K_j X K_j^\dagger \, .
\end{equation} 
By construction we have $\Gamma[(\cE^+ - \cE^-)\otimes \cF] = \cG$.
The maps $\cE^\pm$ are clearly completely positive, so it remains to show that $\cE^+ + \cE^-$ is trace-nonincreasing. For this purpose let $\{\ket{\lambda}\}$ be an orthonormal eigenbasis of $\sum_j \Lambda_j$. For any $\sigma \in \St(A)$ we compute 
\begin{align}
    \tr[(\cE^+ + \cE^-)(\sigma)] = \tr[\sum_j K_j \sigma K_j^\dagger] = \tr[\sum_j \Lambda_j \sigma] = \sum_\lambda \bra{\lambda}\sum_j \Lambda_j \sigma \ket{\lambda} \leq \sum_\lambda \bra{\lambda} \sigma \ket{\lambda} = \tr[\sigma] \, ,
\end{align}
where the inequality follows from the assumption $\sum_j \Lambda_j \leq \id$.

It remains to prove~\cref{eq_second_ds} i.e.\ every Hermitian operator $O$ whose operator norm is bounded by $1$ can be written as $\sum_j b_j \Lambda_j$ like in~\cref{eq_first_ds} and vice versa.
    Clearly the linear map $\sum_j b_j \Lambda_j$ is Hermitian and for every normalized eigenvector $\ket{\lambda}$ to the eigenvalue $\lambda$ we have
    \begin{equation}
        |\lambda| = |\bra{\lambda} \sum_j b_j \Lambda_j \ket{\lambda} | \leq  \sum_{j}| \bra{\lambda} \Lambda_j \ket{\lambda} | = \bra{\lambda} \sum_{j} \Lambda_j \ket{\lambda} \leq 1 \, .
    \end{equation}
    Thus, it holds $\| \sum_j b_j \Lambda_j \| \leq 1$. 
    On the other hand, we can expand an operator $O$ in terms of its spectral decomposition $O = \sum_{\lambda} \lambda \Pi_{\lambda}$ for some orthogonal projections $\{\Pi_{\lambda}\}_{\lambda}$. The bound $\| O \| \leq 1$ implies that for all eigenvalues we have $|\lambda | \leq 1$. Define $\Lambda_{\lambda} = |\lambda| \Pi_{\lambda}$ and the coefficient $b_{\lambda} = \mathrm{sgn}(\lambda)$. Hence, we have $\sum_{\lambda} \Lambda_{\lambda} \leq \id$.
\end{proof}


We next show that $\gamma_{\noCC}$ and $\gamma_{\CC}$ are invariant under unitaries.
The way we introduced wire cutting above, the channel from $A\to B$ does not necessarily have to be the identity---it could already involve a gate. Hence it is natural to ask if (vertically) cutting a gate could be better than cutting the wire beforehand and doing gate locally. Following lemma gives a negative answer to this question.
\begin{proposition} Let $S \in \{\mathrm{LO}(A,B), \mathrm{LOCC}(A,B) \}$, $\cF \in \TPCP(A)$ and let $\cU\in \mathrm{L}(\mathrm{L}(A))$ be a unitary channel. Then
    \begin{equation}
         \gamma_{\Gamma[S]}(\cU \circ \mathcal{F}) = \gamma_{\Gamma[S]}(\mathcal{F}) = \gamma_{\Gamma[S]}( \mathcal{F} \circ \cU) \, .
    \end{equation}
\end{proposition}
\begin{proof}
    We only show the first equality of the proposition, the second follows analogously. We have
    \begin{align}
        \gamma_{\Gamma[S]}(\mathcal{F}) &= \min \{\gamma_S (\Phi) : \Gamma(\Phi)=\mathcal{F}\} \\
        &= \min \{\gamma_S ((\id\otimes\mathcal{U})\circ\Phi) : \Gamma(\Phi)=\mathcal{F}\} \\
        &= \min \{\gamma_S ((\id\otimes\mathcal{U})\circ\Phi) : \Gamma((\id\otimes\mathcal{U})\circ\Phi)=\mathcal{U}\circ\mathcal{F}\} \\
        &= \min \{\gamma_S (\tilde{\Phi}) : \Gamma(\tilde{\Phi})=\mathcal{U}\circ\mathcal{F}\} \\
        &= \gamma_{\Gamma[S]}(\mathcal{U}\circ\mathcal{F}) \, .
    \end{align}
    The first and last step are direct applications of~\cref{same gamma factors}.
    The second step follows from the invariance of the $\gamma$-factor for $\LO$ and $\LOCC$ under local unitaries~\cite{PS22}[Lemma 2.3].
    The third step follows due to basic properties of the the partial trace, which imply that for any $\Phi$ one has  $\Gamma(\Phi)=\mathcal{F}$ if and only if $\Gamma((\id\otimes\mathcal{U})\circ\Phi)=\mathcal{U}\circ\mathcal{F}$.
    The fourth step follows from the substitution $\tilde{\Phi}:=(\id\otimes\mathcal{U})\circ\Phi$.
\end{proof}

\section{Optimal wire cutting}

\subsection{Parallel cuts without classical communication}
Following the discussion from the previous section, the optimal sampling overhead of cutting a single wire without using classical communication is precisely characterized by $\gamma_{\noCC}(\cI_{A\rightarrow B})$ where $A$ and $B$ are one-qubit systems.
If we want to instead consider $n\geq 1$ parallel wires that are to be cut at once, then one has to simply replace $A$ and $B$ by $n$-qubit systems $\tilde{A}:=A_1\otimes A_2\otimes\dots\otimes A_n$ and $\tilde{B}:=B_1\otimes\dots\otimes B_n$.
The sampling overhead in this case is then given by $\gamma_{\noCC}(\cI_{\tilde{A}\rightarrow \tilde{B}})$.
The case of a single wire cut is thus clearly the special case where $n=1$.

The following proposition gives us an explicit formula for the sampling overhead.
\begin{proposition} \label{prop:noCC_parallel} 
Let $n \in \mathbb{N}$ and let $\tilde{A},\tilde{B}$ be $n$-qubit systems. Then
     \begin{equation}
         \gamma_{\noCC(\tilde{A},\tilde{B})}(\cI_{\tilde{A}\rightarrow \tilde{B}}) = 4^n \, .
     \end{equation}
\end{proposition}
This result asserts that the $\gamma$-factor of a wire cut without classical communication behaves multiplicatively under the tensor product.
More concretely, this means that there is no advantage in jointly cutting multiple wires at once when no classical communication is available.\footnote{This is a stark contrast to the setting with classical communication that will be explored in~\cref{sec_cuttingLOCC}.} So finding an optimal QPD for a single wire cut (i.e. a QPD with a $\gamma$-factor of $4$) and then applying that QPD separately for each of the $n$ wire cuts is already optimal. This also implies that the cost for cutting $n$ wires at arbitrary positions in the circuit has an optimal sampling overhead of $4^{2n}$. Hence, \cref{prop:noCC_parallel} fully answers the question how to do optimal wire cuts without classical communication as summarized by~\cref{tab_results}.
We note that the explicit QPD of a single wire cut that achieves the $\gamma$-factor of $4$ is given in~\cref{app_harrow}.

\begin{proof}
An explicit QPD for the identity channel into the set $\noCC$ with one-norm of the coefficients of value $4$ can be derived using techniques from~\cite{PHOW20} and is explicitly given in~\cref{app_harrow}.
This implies $\gamma_{\noCC}(\cI_{\tilde{A}\rightarrow \tilde{B}}) \leq 4^n$ and thus only the converse statement remains to be shown. 

A decomposition for the identity single qubit channel can be written in the form
    \begin{equation} \label{Id channel}
        \cI_{\tilde{A}\rightarrow \tilde{B}} = \sum_{i} a_i \mathcal{E}_i \qquad \textnormal{where} \qquad  \mathcal{E}_i \in \noCC(\tilde{A},\tilde{B}) \, .
    \end{equation}
We denote the generalized $n$-qubit Pauli basis by $\{\id,X,Y,Z\}^{\otimes n}=\{\sigma_1,\sigma_2,\dots,\sigma_{4^{n}}\}$.
Taking the trace of~\cref{Id channel} on both sides in the space $\mathrm{\mathrm{L}(\mathrm{L}(}\mathbb{C}^2))$ yields
\begin{align} \label{eq_step_ds1}
    4^n = \tr[\sum_{i} a_i \mathcal{E}_i]
    = \tr[\sum_{i} a_i \tr[\cdot O_i]\rho_i]
    &= \sum_{i} \frac{a_i}{2^n}\sum_{k=1}^{4^n} \tr[O_i \sigma_k] \tr[\sigma_k \rho_i] \\
    &= \sum_{i,k} \frac{a_i}{2^n} \tr[(O_i \otimes \rho_i)(\sigma_k \otimes \sigma_k)] \, ,
\end{align}
where the second step uses the explicit form of channels $\cE_i \in \noCC$ as shown in~\cref{lem:equivalent_set}.
The third step follows from the fact that the trace of the superoperator $X\mapsto \tr[X\otimes O_i]\rho_i$ is given by $\frac{1}{2^n}\sum_k\tr[O_i\sigma_k]\tr[\sigma_k\rho_i]$.
This can be seen for example by considering its Pauli transfer matrix representation, which has entries
\begin{align}
    M_{\alpha\beta} = \frac{1}{2^n}\tr[\sigma_{\alpha} \tr[\sigma_{\beta}O_i]\rho_i]  = \frac{1}{2^n}\tr[\sigma_{\alpha}\rho_i]\tr[\sigma_{\beta}O_i] \, .
\end{align}
The final step in~\cref{eq_step_ds1} uses that the trace is multiplicative under the tensor product.
Now, we make use of several properties of the $\SWAP$-operator. Recall that if $\{\ket{i}\}_i$ is an orthonormal basis of $\tilde{A}$ and $\tilde{B}$ respectively, we define the $\SWAP$ operator acting between the two systems as $W = \sum_{i,j} \ketbra{i}{j}_{\tilde{A}}\otimes\ketbra{j}{i}_{\tilde{B}}$. 
It is a well known fact that the $\SWAP$-operator $W$ between two $n$-qubit systems can be also written as 
\begin{equation}
    W = \frac{1}{2^n} \sum_{k=1}^{4^n} \sigma_k \otimes \sigma_k \, .
\end{equation}
Hence we can further simplify~\cref{eq_step_ds1} to 
\begin{align}
   4^n = \sum_{i} a_i\tr[(O_i \otimes \rho_i) W] 
     = \sum_{i} a_i \tr[O_i \rho_i]
    \leq \sum_{i} |a_i| | \tr[O_i \rho_i]| 
    \leq \sum_{i} |a_i| \, ,
\end{align}
where the second step uses the SWAP-trick~\cite{carlen_book,ando79}, namely $\mathrm{tr}[(O_{\tilde{A}}\otimes O_{\tilde{B}}) W] = \mathrm{tr}[O_{\tilde{A}} O_{\tilde{B}}]$, where $O_{\tilde{A}}$ and $O_{\tilde{B}}$ are operators acting on $\tilde{A}$ and $\tilde{B}$, respectively. The last inequality follows from Hölder's inequality for matrices (see e.g.~\cite[Proposition~2.5]{Sutter_book}).
This shows $\gamma_{\noCC}(\cI_{\tilde{A}\rightarrow \tilde{B}}) \geq 4^n$.
\end{proof}

\subsection{Parallel cuts with classical communication} \label{sec_cuttingLOCC}
In this section we propose an $\LOCC$ protocol for wire-cutting which is based on the quantum teleportation protocol~\cite{teleporation93}.
Similarly to gate cutting of $\CNOT$ gates in~\cite{PS22} the main idea is to relate wire cutting to quasiprobabilistic simulation of shared Bell pairs.
In~\cref{prop:optimal_CC} we will show that this protocol is optimal in the sense that it achieves the $\gamma$-factor (which is the smallest possible sampling overhead).

\begin{figure}[!htb]
     \centering
    \begin{tikzpicture}
        \draw[thick] (0,0) -- (2,0);
        \draw[thick] (2.5,0) -- (3.5,0);
        \draw[thick] (0,-1) -- (3.5,-1);
        \draw[thick] (0,-2) -- (4.5, -2);
        \draw[thick] (5, -2)-- (6, -2);
        \draw[thick] (6.5,-2) -- (7.5, -2);
        \cnott{1}{0}{-1}{thick};
        \draw[thick] (2,0.25) --(2.5, 0.25) -- (2.5, -0.25) -- (2, -0.25) -- (2, 0.25);
        \draw[thick] (3.5, 0.25) --(4, 0.25) -- (4, -0.25) -- (3.5, -0.25) -- (3.5, 0.25);
        \draw[thick] (3.5, -1.25) --(4, -1.25) -- (4, -0.75) -- (3.5, -0.75) -- (3.5, -1.25);
        \draw[thick] (4.5, -2.25)--(5, -2.25)--(5, -1.75)--(4.5, -1.75)--(4.5,-2.25);
        \draw[thick] (6, -2.25)--(6.5, -2.25)--(6.5, -1.75)--(6, -1.75)--(6,-2.25);
        \draw[thick, snake it] (0,-1) -- (0,-2);
        \draw[thick, double] (4, -1) -- (4.75, -1) -- (4.75, -1.75);
        \draw[thick, double] (4, 0) -- (6.25, 0) -- (6.25, -1.75);
        \node at (0, -1) [circle,fill,inner sep=1pt]{};
        \node at (0, -2) [circle,fill,inner sep=1pt]{};
        \node at (4.75, -1) [circle,fill,inner sep=1pt]{};
        \node at (6.25, 0) [circle,fill,inner sep=1pt]{};
        \node at (2.25,0) {$H$};
        \node at (4.375, -0.75) {$b$};
        \node at (5.125, 0.25) {$a$};
        \node at (6.25, -2) {$Z^a$};
        \node at (4.75, -2) {$X^b$};
        \draw (3.95,-1.13) arc (0:180:0.2);
        \draw[->] (3.72,-1.1) -- (3.9,-0.82);
        \draw (3.95,-1.13+1) arc (0:180:0.2);
        \draw[->] (3.72,-1.1+1) -- (3.9,-0.82+1);        
        \node at (-0.5, 0) {$\ket{\psi}$};
        \node at (8, -2) {$\ket{\psi}$};
        \node at (-0.5, -1.5) {$\ket{\Psi}$};
    \end{tikzpicture}
\caption{Circuit performing a teleportation of the state $\ket{\psi}$ using an ebit $\ket{\Psi}$ and LOCC.}
\label{fig_state_teleporation}
\end{figure}
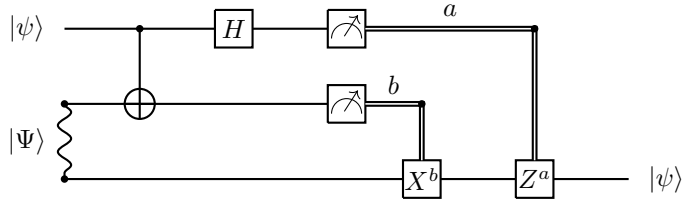

The state teleportation protocol is a classic result in quantum information theory that allows two parties that own a pre-shared Bell pair to transfer one qubit of information by just performing $\LOCC$ operations.
The protocol is summarized in~\cref{fig_state_teleporation} and reduces the problem of quasiprobabilistic wire cutting to the problem of quasiprobabilistically simulating the preparation of a shared Bell pair.
Clearly, if $n\geq 1$ wires are to be cut in parallel at once, this can be achieved by first simulating $n$ shared Bell pairs and applying the quantum teleportation protocol $n$ times.
Luckily, as we discussed in~\cref{sec:qp_gate_cutting}, the optimal QPD of shared Bell pairs and its associated $\gamma$-factor are well understood and by~\cref{eq:gamma_n_bell_pairs} we see that our protocol provides an upper bound of $2^{n+1}-1$ for the $\gamma$-factor of $n$ wire cuts.

Following result asserts that the gate teleportation protocol is indeed optimal:
\begin{proposition} \label{prop:optimal_CC}
    Let $n \in \mathbb{N}$ and let $\tilde{A},\tilde{B}$ be $n$-qubit systems. Then
     \begin{equation}
         \gamma_{\CC(\tilde{A},\tilde{B})}(\cI_{\tilde{A}\rightarrow\tilde{B}}) = 2^{n+1} - 1 \, .
     \end{equation}
\end{proposition}
We stress here that compared to the case without classical communication summarized in~\cref{prop:noCC_parallel}, the $\gamma$-factor of a wire cut features a strictly submultiplicative behavior under the tensor product when classical communication is allowed.
This has major implications for the practical realization of wire cutting of parallel wires: Optimally cutting multiple wires at once exhibits a smaller overhead than optimally cutting them all individually.
Another interesting observation is that is that even in the case of a single wire cut (i.e.~$n=1$) classical communication reduces the $\gamma$-factor from $4$ to $3$. This is in stark contrast to gate cutting where for a single gate instance there is no difference between the $\gamma$-factor for $\LO$ and $\LOCC$ as discussed in~\cref{sec:qp_gate_cutting}.

\begin{proof}
We first show the statement for the case $n=1$. 
Using a QPD of an ebit (see~\cref{eq:gamma_n_bell_pairs}), we can achieve a wire cut with an overhead of $\kappa = 3$ exploiting quantum teleportation as an identity channel between two parties of a bipartite system. This shows that $\gamma_{\CC(\tilde{A},\tilde{B})}(\cI_{\tilde{A} \to \tilde{B}}) \leq 3$.

On the other hand, we one can clearly create an ebit by performing a wire cut.
More mathematically, consider three qubit systems $A$, $B$ and $C$ and assume that we have a QPD of the identity channel
\begin{equation} \label{Id channel}
    \cI_{C\rightarrow B} = \sum_{i} a_i \mathcal{E}_i \qquad \textnormal{where} \qquad  \mathcal{E}_i \in \CC(C,B) \, .
\end{equation}
For every $i$, we choose a map $\mathcal{F}_i\in \LOCC(C,B)$ s.t. $\Gamma[\mathcal{F}_i] = \mathcal{E}_i$.

Clearly, a Bell pair $\ket{\Psi}_{AB}$ can be realized by taking a Bell pair $\ket{\Psi}_{AC}$ and applying the identity map $\cI_{C\rightarrow B}$, which implies
\begin{align}
    \proj{\Psi}_{AB} &= (\cI_A\otimes \cI_{C\rightarrow B})(\proj{\Psi}_{AC}) \\
    &= \sum\limits_i a_i (\cI_A\otimes \mathcal{E}_i)(\proj{\Psi}_{AC}) \\
    &= \sum\limits_i a_i \tr_C[(\cI_A\otimes \mathcal{F}_i)(\proj{\Psi}_{AC}\otimes\proj{0}_B)] \, .
\end{align}
Since any LOCC operation (here across $A\otimes C$ and $B$) can only generate separable states, we can write the resulting state as a probabilistic mixture of product states:
\begin{equation}
    (\cI_A\otimes \mathcal{F}_i)(\proj{\Psi}_{AC}\otimes\proj{0}_B)=\sum_j p_{i,j} \tau_{i,j,AC}\otimes \sigma_{i,j,B}
\end{equation}
where $\tau_{i,j,AC},\sigma_{i,j,B}$ are Hermitian and $\|\tau_{i,j,AC}\|_1,\|\sigma_{i,j,B}\|_1\leq 1$.
This clearly gives us a QPD for a Bell pair across $A$ and $B$:
\begin{align}
    \proj{\Psi}_{AB} = \sum\limits_{i,j} a_ip_{i,j} \tr_C[\tau_{i,j,AC}]\otimes \sigma_{i,j,B}  \, .
\end{align}
The one-norm of the coefficients of this QPD are identical, since $\sum_{i,j} |a_i p_{i,j}| = \sum_i |a_i| \sum_j p_{i,j} = \sum_i |a_i|$.
The operators $\tr_C[\tau_{i,j,AC}]$ and $\sigma_{i,j,B}$ can be separate into a positive and a negative parts, which similarly does not impact the one-norm of the quasiprobability coefficients.
This shows $\gamma_{\CC}(\cI_{A}) \geq \gamma_{\LOCC}(\ket{\Psi}) = 3$.

The proof for $n=1$ above can straightforwadly be lifted to general $n>1$.
Clearly, $2^{n+1}-1$ is achieved by the protocol explained in the main text that realizes the $n$ wire cuts using $n$ instances of state teleportation for which we jointly prepare the $n$ ebits.
For the converse, one simply concatenates the $n=1$ argument $n$ times.
$n$ ebits can be realized through $n$ wire cuts, so any QPD of $n$ wire cuts automatically implies a QPD for $n$ ebits with the same one norm of the coefficients.
\end{proof}

\subsection{Arbitrary cuts with classical communication} \label{sec_arbitrary_cuts_LOCC}
\cref{prop:optimal_CC} is stated in terms of parallel wire cuts.
However, the quantum teleportation-based protocol that achieves this bound works in a more general setting: The Bell pairs that are consumed by individual wire cuts do not need to be simulated just before the cut itself occurs, instead, they can be simulated ahead of time and stored in ancilla memory qubits.
In principle, all wire cuts of a circuit could be generated using gate teleportation and the involved Bell pairs could all be simulated at the very beginning of the circuit.
Since the Bell pairs are generated at the same time (even though the wire cuts themselves occur at different time slices in the circuit), one can still profit from the submultiplicativity behavior of the $\gamma$-factor of Bell pairs and achieve a $\gamma$-factor of $2^{n+1}-1$ for $n$ wire cuts.
This is in stark contrast with the method in~\cite{xandadu_22} that can only exploit submultiplicativity when the cuts truly occur in parallel (i.e.~within the same time slice).

Of course, if one where to generate all $n$ Bell pairs at the very beginning of the circuit, the size of this additional memory required by the technique would grow linearly with the number $n$ of wire cuts.
This is inconvenient, as the limited number of qubits precisely is the motivation for circuit knitting in the first place.
For practical purposes, it is more useful to generate a fixed number $k\leq n$ of Bell pairs at a time and then, once all $k$ Bell pairs have been consumed by quantum teleportation protocols, reuse these $2k$ qubits to generate new Bell pairs.
Choosing the size $k$ of this \emph{entanglement factory} comes with a tradeoff: Choosing $k$ smaller results in a reduced memory footprint of the method and increasing $k$ decreases the effective $\gamma$-factor per wire cut, which is given by $(2^{k+1}-1)^{1/k}$.

It should be noted that if this teleportation-based wire cutting technique is combined together with the gate teleportation-based gate cutting technique of~\cite{PS22}, then both methods can share the same entanglement factory.

In this section we have seen an algorithm to cut $n$ wires at arbitrary positions in a circuit with a sampling overhead $(2^{n+1} -1)^2$.
Since the task of cutting $n$ wires in parallel is clearly a special case of cutting $n$ wires at arbitrary positions, the optimal sampling overhead of the latter must in general be at least as large as the sampling overhead of the former.
By~\cref{prop:optimal_CC}, this implies that our method is also optimal for $n$ arbitrary cuts as noted in~\Cref{tab_results}.

\section{Discussion}\label{sec_discussion}
Circuit knitting consists of cutting gates and wires. Together with the gate cutting techniques presented in~\cite{PS22}, this work therefore complements the understanding of optimal circuit cutting within the quasiproability simulation framework.

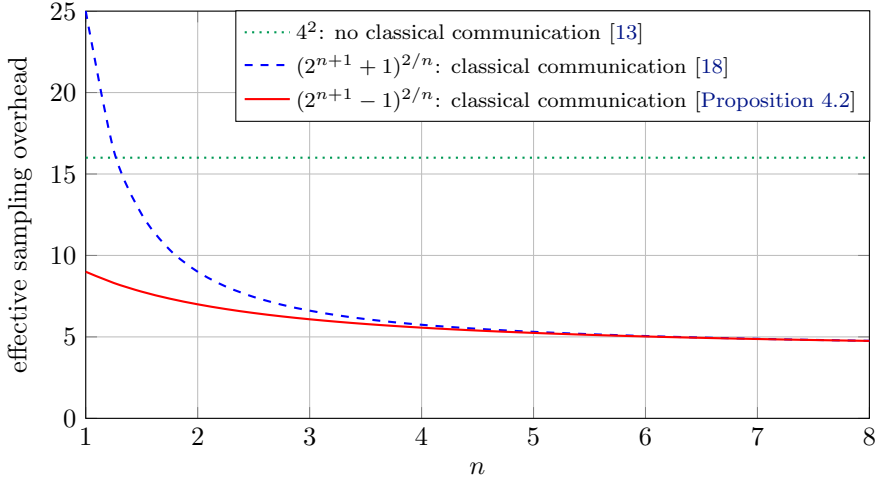
\begin{figure}[!htb]
\centering

  \begin{tikzpicture}
	\begin{axis}[
		height=7.0cm,
		width=12.0cm,
		grid=major,
		xlabel=$n$,
		ylabel=effective sampling overhead,
		xmin=1,
		xmax=8,
		ymax=25,
		ymin=0,
	     xtick={1,2,3,4,5,6,7,8},
         ytick={25,20,15,10,5,0},
		legend style={at={(0.596,1.002)},anchor=north,legend cell align=left,font=\footnotesize} 
	]
	\addplot[thick, dotted, ForestGreen,smooth,name path=f] coordinates {
(1,16) (8,16)};
	\addlegendentry{$4^2$: no classical communication~\cite{PHOW20}}
	
	\addplot[thick, dashed,blue,smooth,name path=f] coordinates {
(1.,25.) (1.25,16.4548) (1.5,12.5225) (1.75,10.3486) (2.,9.) (2.25,8.09516) (2.5,7.45266) (2.75,6.97646) (3.,6.61149) (3.25,6.3241) (3.5,6.09269) (3.75,5.90284) (4.,5.74456) (4.25,5.61077) (4.5,5.49629) (4.75,5.39729) (5.,5.31087) (5.25,5.23477) (5.5,5.16726) (5.75,5.10697) (6.,5.05277) (6.25,5.0038) (6.5,4.9593) (6.75,4.9187) (7.,4.88149) (7.25,4.84725) (7.5,4.81564) (7.75,4.78636) (8.,4.75915)};
	\addlegendentry{$(2^{n+1}+1)^{2/n}$: classical communication~\cite{xandadu_22}}

	\addplot[thick,red,smooth,name path=f] coordinates {
	(1.,9.) (1.25,8.31217) (1.5,7.77661) (1.75,7.34882) (2.,7.) (2.25,6.71073) (2.5,6.46743) (2.75,6.26033) (3.,6.0822) (3.25,5.92762) (3.5,5.79242) (3.75,5.67332) (4.,5.56776) (4.25,5.47367) (4.5,5.38937) (4.75,5.31349) (5.,5.24489) (5.25,5.18263) (5.5,5.12591) (5.75,5.07407) (6.,5.02653) (6.25,4.9828) (6.5,4.94247) (6.75,4.90518) (7.,4.8706) (7.25,4.83848) (7.5,4.80855) (7.75,4.78062) (8.,4.7545)};
	\addlegendentry{$(2^{n+1}-1)^{2/n}$: classical communication [\cref{prop:optimal_CC}]}		
	\end{axis} 
\end{tikzpicture}
\caption{Sampling overhead per wire cut for different scenarios. The x-axis denotes the number of wire cuts.}
\label{fig_tradeoff}
\end{figure}

One of the key results of this work is that without classical communication the optimal sampling overhead for cutting $n$ wires is multiplicative. In words, this means that the optimal cost for cutting $n$ wires equals the cost for cutting a single wire $n$-times (see~\cref{tab_results}). This changes drastically if we allow for classical communication. In theses settings the $\gamma$-factor is strictly submultiplicative, i.e., the cost for cutting $n$ wires is strictly smaller than the cost for cutting $n$-times a single wire (see~\cref{tab_results}).
\Cref{fig_tradeoff} shows the cost for a single wire cut in the case where we can do several wire cuts together (which we call \emph{effective sampling overhead}). The mentioned submultiplicativity property manifests itself by the decrease of the curves.
We see that there is a provable improvement of the sampling overhead if we allow for classical communication (even in case $n=1$).

Last, we would like to emphasize that our protocol for cutting $n$ wires at arbitrary positions in the circuit, which achieves the optimal sampling overhead, requires $n$ ancilla qubits.
However, in practice we keep the number of required ancillas constant by resusing the ancilla bits, as discussed in~\cref{sec_cuttingLOCC}.
It is left as an open question for future research if the optimal sampling overhead for wire cutting $n$ qubits (either in parallel or at arbitrary locations) is possible without the need for additional ancilla qubits.\footnote{Note that the protocol presented in~\cite{xandadu_22} for parallel wire cutting does not require any ancilla qubits, however also does not achieve the optimal sampling overhead.}

\begin{remark}[Emulation of one-way classical communication]
Interestingly, the wire cutting protocol presented in~\cref{sec_arbitrary_cuts_LOCC} does not actually require bi-directional classical communication between the two sub-circuits.
Rather, there is only one-way communication from the system sending the qubit to the system receiving the qubit.
For this reason, the classical communication can be emulated in some cases, foregoing the requirement of physically having two separate quantum computers that exchange classical information in real time.

More precisely, when a circuit is cut such that no wire is cut more than once (see for example~\cref{fig_parallel}), then the cut effectively separates the circuit into two temporally separated parts.
This makes it possible to sequentially execute the two circuits one after another and realizing the classical communication by storing the outcomes of the first circuit and then using these outcomes when executing the second circuit.
In fact, the two subcircuits can even be ran on the same physical quantum computer.

However, once there exists a wire that is cut more than once (see for example~\cref{fig_arbitrary}), then this is not possible anymore as the wire cutting involves classical communication in both directions.
\end{remark}

\begin{example}[Gate cuts vs wire cuts]
To exemplify the usefulness of our results on the circuits in~\cref{fig_both_cuts}, we can use~\cref{tab_results} to quantify the costs for the different possible cuts depicted in the figure.
\Cref{tab_example} summarizes the resulting sampling overheads.\footnote{Recall that the optimal cost for cutting a single $\CNOT$ gate is $3^2$ as shown in~\cref{eq_CNOT_cost}.} 
As we see the optimal cut for~\Cref{fig_line_cit} and~\Cref{fig_gate_cut} are two wire cuts and one gate cut, respectively.
\begin{table}[!htb]
\centering
\bgroup
\def\arraystretch{1.5}
  \begin{tabular}{V{2.5}lV{2.5}c|cV{2.5}c|cV{2.5}}
  \clineB{2-5}{2.5}  
\multicolumn{1}{cV{2.5}}{}&
      \multicolumn{2}{cV{2.5}}{\Cref{fig_line_cit}} &
      \multicolumn{2}{cV{2.5}}{\Cref{fig_gate_cut}} \\ \cline{2-5}
\multicolumn{1}{cV{2.5}}{}& 2 lines [\Cref{tab_results}] & 3 $\CNOT$~\cite{PS22} & \!2 lines [\Cref{tab_results}]\!   & \!1 $\CNOT$~\cite{PS22}\!  \\ 
\clineB{1-5}{2.5}
\!\!Without classical comm.  & $(4^2)^2 = 16^2$ & $(3^3)^2 = 27^2$  & $(4^2)^2 = 16^2$ & $3^2$\\
\clineB{1-5}{2.5}
\!\!With classical comm. & $(2^{2+1}-1)^2 = 7^2$ & $(2^{3+1}-1)^2 = 15^2$ & $(2^{2+1}-1)^2 = 7^2$ & $3^2$ \\
\clineB{1-5}{2.5}
  \end{tabular}
\egroup
		\caption{Optimal sampling overhead for the wire and gate cuts presented in~\cref{fig_both_cuts} with and without classical communication between the subcircuits.}
		\label{tab_example}
	\end{table}
\end{example}

\paragraph{Acknowledgements}
We thank Stefan W\"orner and Sebastian Brandhofer for fruitful discussions on wire cutting.
CP achknowledges support by the Swiss National Science Foundation, through the National Center of Competence in Research ``Quantum Science and Technology'' (QSIT) and through grant number 20QT21\_187724.

\appendix
\section{Quasiprobability decompositions in existing wire cutting methods} \label{app_QPD_methods}
In this appendix we summarize the quasiprobability decompositions of the identity gate used in previous works on wire cutting~\cite{PHOW20,xandadu_22}.
This allows us to show how these methods are fully described by the formalism described in~\cref{sec:reduction_wire_to_gate}.

\subsection{Wire cutting without classical communication introduced in~\cite{PHOW20}} \label{app_harrow}
The wire cutting method presented in~\cite{PHOW20} is based on the following decomposition of the identity channel 
\begin{align} \label{eq_QPD_Harrow}
\cI(\cdot) = \sum_{i=1}^8 a_i \cG_i(\cdot) \quad \textnormal{where} \quad \mathcal{G}_i(A) = \tr[O_iA] \rho_i \, ,
\end{align}
and $A \in\mathrm{L}(\mathbb{C}^2) $ and $\{O_i,\rho_i,a_i\}$ given in~\cref{table noCC QPD}.
The decomposition relies on the fact that the Pauli matrices $\id, X, Y, Z$ form a basis of $\mathrm{L}(\mathbb{C}^2)$. 
By~\cref{lem:equivalent_set} the $\mathcal{G}_i$ are clearly elements of $\noCC$ and thus do not involve any classical communication.
We see that the sampling overhead is given by $(\sum_i a_i)^2 = 4^2$. \cref{prop:noCC_parallel} proves that this QPD is optimal.

\begin{table}[h] 
\centering
\begin{tabular}{c|c|c}
   $O_i$ & $\rho_i$  & $a_i$ \\ \hline
    $\id$ & $\proj{0}$ & $+1/2$ \\ \hline
    $\id$ & $\proj{1}$ & $+1/2$ \\ \hline
    $X$ & $\proj{+}$ & $+1/2$ \\ \hline
    $X$ & $\proj{-}$ & $-1/2$ \\ \hline
    $Y$ & $\proj{i}$ & $+1/2$ \\ \hline
    $Y$ & $\ket{-i}\bra{-i}$ & $-1/2$ \\ \hline
    $Z$ & $\proj{0}$ & $+1/2$ \\ \hline
    $Z$ & $\proj{1}$ & $-1/2$ 
\end{tabular}
\caption{Operators, states and weights required to implement a single wire cut using~\cref{eq_QPD_Harrow}.}
\label{table noCC QPD}
\end{table}

\subsection{Wire cutting with classical communication introduced in \cite{xandadu_22}}
Let us consider an $n$ qubit system and let $d = 2^n$ be the dimension of the Hilbert space. Let $\{\ket{j} \}_{0 \leq j \leq d-1}$ be the standard basis of the Hilbert space and $\{V_j\}_{0 \leq j \leq d-1} \subset \mathcal{U}(\mathbb{C}^d)$ be  unitaries characterized by the condition $V_j \ket{0} = \ket{j}$. The authors of~\cite{xandadu_22} realize $n$ parallel wire cuts through the decomposition 
\begin{align}
\cI(\cdot) = (d+1) \Psi_0(\cdot) - d\Psi_1(\cdot) \, ,
\end{align}
where $\Psi_0(A) = \mathbb{E}_U[ \sum_j \bra{j} U^\dagger A U \ket{j} U \proj{j} U^\dagger]$ corresponds to a measure-and-prepare channel with respect to the standard basis transformed by the Haar-random unitary $U$. 
In order to implement this channel practically, one can pick a (finite) unitary 2-design $\mathrm{T} \subset \mathrm{U}(\mathbb{C}^d)$. 
In addition, $\Psi_1(A) = \tr[A]\id /d $ is the fully depolarizing channel.

To see that both $\Psi_0$ and $\Psi_1$ are elements of $\CC$, we note that $\Psi_0=\Gamma[\mathbb{E}_U\mathcal{E}_U]$ and $\Psi_1=\Gamma[\cI\otimes\Psi_1]$ where
\begin{align}
    \mathcal{E}_U := \sum_{j} \mathrm{Ad}[U \proj{j} U^\dagger] \otimes \mathrm{Ad}[UV_j]
\end{align}
for $\mathrm{Ad}[A](X):= AXA^{\dagger}$.
The sampling overhead of this QPD is clearly $((d+1) + d)^2=(2d+1)^2=(2^{n+1}+1)^2$.

\bibliographystyle{arxiv_no_month}
\bibliography{bibliofile}

\begin{thebibliography}{10}

\bibitem{IBM_22}
S.~Bravyi, O.~Dial, J.~M. Gambetta, D.~Gil, and Z.~Nazario.
\newblock The future of quantum computing with superconducting qubits.
\newblock {\em Journal of Applied Physics}, 132(16):160902, 2022.
\newblock
  \texttt{\href{http://dx.doi.org/10.1063/5.0082975}{DOI:\,10.1063/5.0082975}}.

\bibitem{TBG17}
K.~Temme, S.~Bravyi, and J.~M. Gambetta.
\newblock Error mitigation for short-depth quantum circuits.
\newblock {\em Phys. Rev. Lett.}, 119:180509, 2017.
\newblock
  \texttt{\href{http://dx.doi.org/10.1103/PhysRevLett.119.180509}{DOI:\,10.1103/PhysRevLett.119.180509}}.

\bibitem{endo18}
S.~Endo, S.~C. Benjamin, and Y.~Li.
\newblock Practical quantum error mitigation for near-future applications.
\newblock {\em Phys. Rev. X}, 8:031027, 2018.
\newblock
  \texttt{\href{http://dx.doi.org/10.1103/PhysRevX.8.031027}{DOI:\,10.1103/PhysRevX.8.031027}}.

\bibitem{kandala19}
A.~Kandala, K.~Temme, A.~D. C{\'o}rcoles, A.~Mezzacapo, J.~M. Chow, and J.~M.
  Gambetta.
\newblock Error mitigation extends the computational reach of a noisy quantum
  processor.
\newblock {\em Nature}, 567(7749):491--495, 2019.
\newblock
  \texttt{\href{http://dx.doi.org/10.1038/s41586-019-1040-7}{DOI:\,10.1038/s41586-019-1040-7}}.

\bibitem{PSBGT21}
C.~Piveteau, D.~Sutter, S.~Bravyi, J.~M. Gambetta, and K.~Temme.
\newblock Error mitigation for universal gates on encoded qubits.
\newblock {\em Phys. Rev. Lett.}, 127:200505, 2021.
\newblock
  \texttt{\href{http://dx.doi.org/10.1103/PhysRevLett.127.200505}{DOI:\,10.1103/PhysRevLett.127.200505}}.

\bibitem{PSW22}
C.~Piveteau, D.~Sutter, and S.~Woerner.
\newblock Quasiprobability decompositions with reduced sampling overhead.
\newblock {\em npj Quantum Information}, 8(1):12, 2022.
\newblock
  \texttt{\href{http://dx.doi.org/10.1038/s41534-022-00517-3}{DOI:\,10.1038/s41534-022-00517-3}}.

\bibitem{PWB15}
H.~Pashayan, J.~J. Wallman, and S.~D. Bartlett.
\newblock Estimating outcome probabilities of quantum circuits using
  quasiprobabilities.
\newblock {\em Phys. Rev. Lett.}, 115:070501, 2015.
\newblock
  \texttt{\href{http://dx.doi.org/10.1103/PhysRevLett.115.070501}{DOI:\,10.1103/PhysRevLett.115.070501}}.

\bibitem{HC17}
M.~Howard and E.~Campbell.
\newblock Application of a resource theory for magic states to fault-tolerant
  quantum computing.
\newblock {\em Phys. Rev. Lett.}, 118:090501, 2017.
\newblock
  \texttt{\href{http://dx.doi.org/10.1103/PhysRevLett.118.090501}{DOI:\,10.1103/PhysRevLett.118.090501}}.

\bibitem{SC19}
J.~R. Seddon and E.~T. Campbell.
\newblock Quantifying magic for multi-qubit operations.
\newblock {\em Proceedings of the Royal Society A: Mathematical, Physical and
  Engineering Sciences}, 475(2227):20190251, 2019.
\newblock
  \texttt{\href{http://dx.doi.org/10.1098/rspa.2019.0251}{DOI:\,10.1098/rspa.2019.0251}}.

\bibitem{HG19}
M.~Heinrich and D.~Gross.
\newblock Robustness of magic and symmetries of the stabiliser polytope.
\newblock {\em {Quantum}}, 3:132, 2019.
\newblock
  \texttt{\href{http://dx.doi.org/10.22331/q-2019-04-08-132}{DOI:\,10.22331/q-2019-04-08-132}}.

\bibitem{SBHYC21}
J.~R. Seddon, B.~Regula, H.~Pashayan, Y.~Ouyang, and E.~T. Campbell.
\newblock Quantifying quantum speedups: Improved classical simulation from
  tighter magic monotones.
\newblock {\em PRX Quantum}, 2:010345, 2021.
\newblock
  \texttt{\href{http://dx.doi.org/10.1103/PRXQuantum.2.010345}{DOI:\,10.1103/PRXQuantum.2.010345}}.

\bibitem{Piv_masterThesis}
C.~Piveteau.
\newblock Advanced methods for quasiprobabilistic quantum error mitigation.
\newblock
  \texttt{\href{http://dx.doi.org/10.3929/ethz-b-000504508}{DOI:\,10.3929/ethz-b-000504508}}.
\newblock Master thesis, ETH Zurich, September 2020.

\bibitem{PHOW20}
T.~Peng, A.~W. Harrow, M.~Ozols, and X.~Wu.
\newblock Simulating large quantum circuits on a small quantum computer.
\newblock {\em Phys. Rev. Lett.}, 125:150504, 2020.
\newblock
  \texttt{\href{http://dx.doi.org/10.1103/PhysRevLett.125.150504}{DOI:\,10.1103/PhysRevLett.125.150504}}.

\bibitem{Mitarai_2021}
K.~Mitarai and K.~Fujii.
\newblock Constructing a virtual two-qubit gate by sampling single-qubit
  operations.
\newblock {\em New Journal of Physics}, 23(2):023021, 2021.
\newblock
  \texttt{\href{http://dx.doi.org/10.1088/1367-2630/abd7bc}{DOI:\,10.1088/1367-2630/abd7bc}}.

\bibitem{MF_21}
K.~Mitarai and K.~Fujii.
\newblock Overhead for simulating a non-local channel with local channels by
  quasiprobability sampling.
\newblock {\em {Quantum}}, 5:388, 2021.
\newblock
  \texttt{\href{http://dx.doi.org/10.22331/q-2021-01-28-388}{DOI:\,10.22331/q-2021-01-28-388}}.

\bibitem{forging22}
A.~Eddins, M.~Motta, T.~P. Gujarati, S.~Bravyi, A.~Mezzacapo, C.~Hadfield, and
  S.~Sheldon.
\newblock Doubling the size of quantum simulators by entanglement forging.
\newblock {\em PRX Quantum}, 3:010309, 2022.
\newblock
  \texttt{\href{http://dx.doi.org/10.1103/PRXQuantum.3.010309}{DOI:\,10.1103/PRXQuantum.3.010309}}.

\bibitem{PS22}
C.~Piveteau and D.~Sutter.
\newblock Circuit knitting with classical communication, 2022.
\newblock
  \texttt{\href{http://dx.doi.org/10.48550/ARXIV.2205.00016}{DOI:\,10.48550/ARXIV.2205.00016}}.

\bibitem{xandadu_22}
A.~Lowe, M.~Medvidović, A.~Hayes, L.~J. O'Riordan, T.~R. Bromley, J.~M.
  Arrazola, and N.~Killoran.
\newblock Fast quantum circuit cutting with randomized measurements, 2022.
\newblock
  \texttt{\href{http://dx.doi.org/10.48550/ARXIV.2207.14734}{DOI:\,10.48550/ARXIV.2207.14734}}.

\bibitem{CLMOW14}
E.~Chitambar, D.~Leung, L.~Man{\v c}inska, M.~Ozols, and A.~Winter.
\newblock Everything you always wanted to know about {LOCC} (but were afraid to
  ask).
\newblock {\em Communications in Mathematical Physics}, 328(1):303--326, 2014.
\newblock
  \texttt{\href{http://dx.doi.org/10.1007/s00220-014-1953-9}{DOI:\,10.1007/s00220-014-1953-9}}.

\bibitem{carlen_book}
E.~Carlen.
\newblock {\em Trace Inequalities and Quantum Entropy: An Introductory Course}.
\newblock Contemporary Mathematics, 2009.
\newblock
  \texttt{\href{http://dx.doi.org/10.1090/conm/529}{DOI:\,10.1090/conm/529}}.

\bibitem{ando79}
T.~Ando.
\newblock Concavity of certain maps on positive definite matrices and
  applications to {H}adamard products.
\newblock {\em Linear Algebra and its Applications}, 26:203--241, 1979.
\newblock
  \texttt{\href{http://dx.doi.org/https://doi.org/10.1016/0024-3795(79)90179-4}{DOI:\,https://doi.org/10.1016/0024-3795(79)90179-4}}.

\bibitem{Sutter_book}
D.~Sutter.
\newblock {\em Approximate Quantum Markov Chains}.
\newblock Springer International Publishing, 2018.
\newblock
  \texttt{\href{http://dx.doi.org/10.1007/978-3-319-78732-9{\_}5}{DOI:\,10.1007/978-3-319-78732-9{\_}5}}.

\bibitem{teleporation93}
C.~H. Bennett, G.~Brassard, C.~Cr\'epeau, R.~Jozsa, A.~Peres, and W.~K.
  Wootters.
\newblock Teleporting an unknown quantum state via dual classical and
  einstein-podolsky-rosen channels.
\newblock {\em Phys. Rev. Lett.}, 70:1895--1899, 1993.
\newblock
  \texttt{\href{http://dx.doi.org/10.1103/PhysRevLett.70.1895}{DOI:\,10.1103/PhysRevLett.70.1895}}.

\end{thebibliography}

\end{document}